\documentclass[
    ,final            
  ]
  {aipproc}

\layoutstyle{6x9}

\newcommand{\be}{\begin{equation}}
\newcommand{\ee}{\end{equation}}
\newcommand{\ba}{\begin{eqnarray}}
\newcommand{\ea}{\end{eqnarray}}
\newcommand{\ci}[1]{\cite{#1}}

\def\vb0{{\bf b}_0}

\def\mev{\,{\rm MeV}}
\def\gev{\,{\rm GeV}}

\def\xbj{x_{\rm Bj}}
\newcommand{\sla}{\hspace*{-0.20cm}/}

\newcommand{\lsim}{\raisebox{-4pt}{$\,\stackrel{\textstyle
                                                         <}{\sim}\,$}}

\newcommand{\req}[1]{(\ref{#1})}
\def\xb{\bar{x}}

\def\={\,=\,}
\def\eps{\epsilon}

\begin{document}

\title{Deeply virtual vector meson electroproduction at small Bjorken-x}


\author{P.\ Kroll}{
address={Fachbereich Physik, Universit\"at Wuppertal, D-42097 Wuppertal, Germany}
}

\begin{abstract}
It is reported on an analysis of vector meson electroproduction at
small Bjorken-$x$ ($\xbj$) within the handbag approach. Using a 
model for the generalized parton distributions (GPDs) and calculating 
the partonic subprocess, electroproduction off gluons, within the 
modified perturbative approach, cross sections and spin density
matrix elements (SDME) are evaluated. The numerical results of this 
analysis agree fairly well with recent HERA data.\\
{\it Contribution to DIS05, Madison (USA)}
\end{abstract}

\maketitle

\begin{figure}[t]
\includegraphics[width=.43\textwidth,bb=112 506 350 715,clip=true]{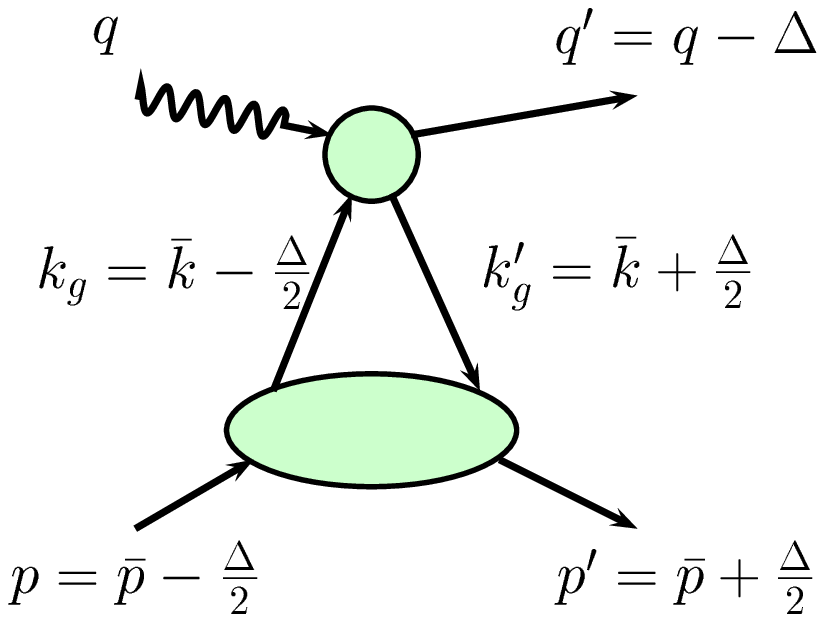}
\includegraphics[width=.43\textwidth,bb=63 355 530 747,clip=true]{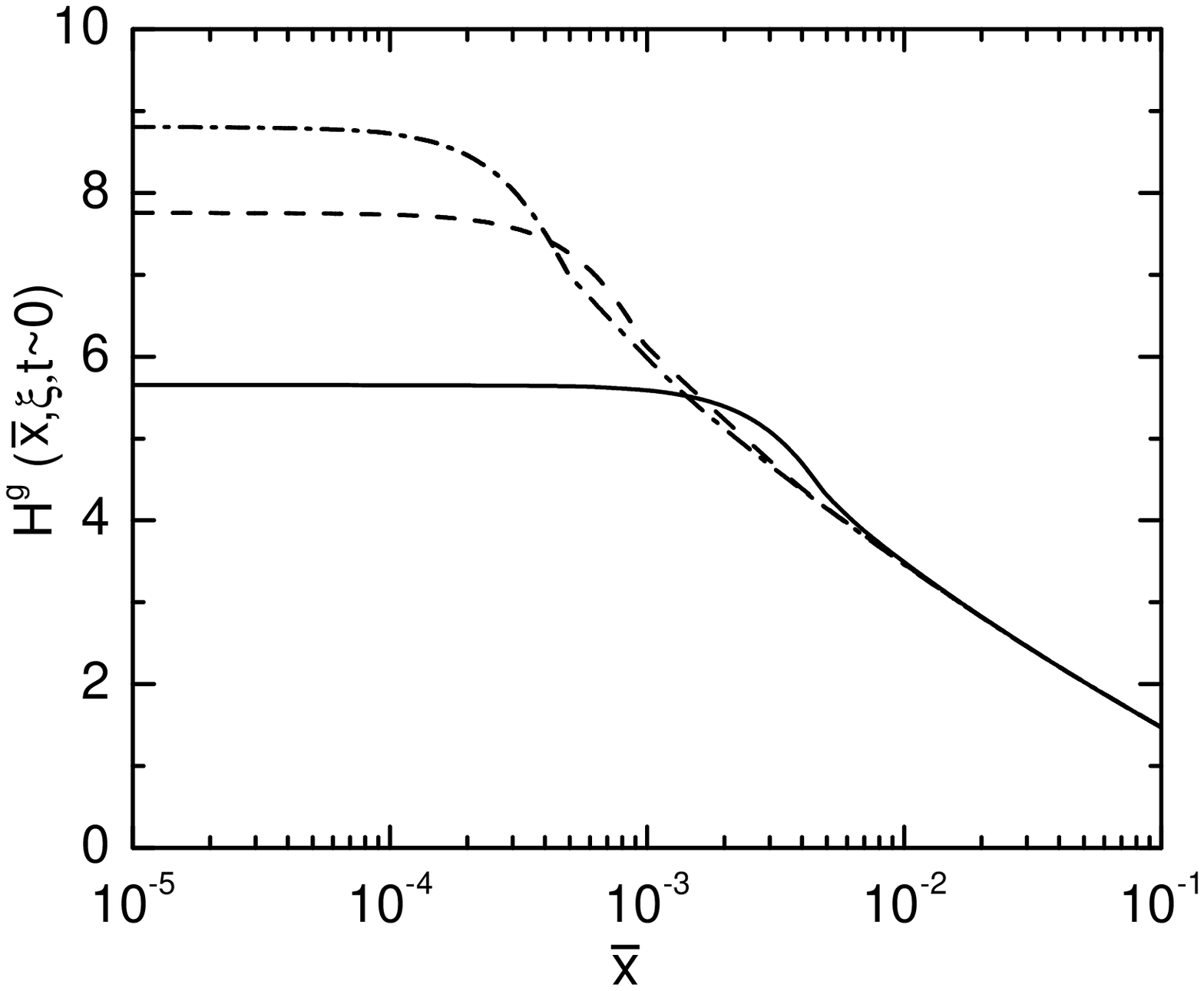}
\caption{Left: The handbag diagram for meson electroproduction off
protons. The large blob represents a GPD while the small one stands
for the subprocess. The momenta of the involved particles are specified.
Right: Model results for the GPD $H^g$ at $t\simeq 0$ and for the case 
$n=1$. The results for $n=2$ are similar. The solid (dashed,
dash-dotted) line represents the GPD at 
$\xi= 5\;(1\,,\; 0.5)\, \cdot 10^{-3}$ and at a scale of $2\,\gev$.}
\label{fig:1}
\end{figure}

It has been shown \ci{rad96} that, at large photon 
virtuality $Q^2$, meson electroproduction factorizes in a partonic 
subprocess, electroproduction off gluons or quarks, 
$\gamma^* g(q)\to V g(q)$, and GPDs, representing soft proton matrix 
elements (see Fig.\ \ref{fig:1}). At small $\xbj$ ($\lsim 10^{-2}$)
the quark subprocesses can be ignored. In the
following I am going to report on an analysis of vector meson
electroproduction within this handbag approach \ci{golo} in the
kinematical regime of large $Q^2$ and large energy $W$ in the photon-proton
c.m.s. but small $\xbj$ and Mandelstam $t$. An exploratory study of
the longitudinal cross section $\sigma_L$ for $\gamma^* p\to Vp$ has 
been performed by Mankiewicz et al. \ci{piller} within this approach. 
Effects of the GPDs have been estimated by Martin et al \ci{martin}.

The structure of the proton is rather complex. In correspondence to
its four form factors there are four gluon GPDs $H^g$, $E^g$,
$\widetilde{H}^g$ and $\widetilde{E}^g$ and four for each quark
flavour. All GPDs are functions of three variables, $t$, skewness
$\xi$ and the average momentum fraction $\xb$, the latter two are 
defined by  
\be 
\xi \= \frac{(p-p')^+}{(p + p')^+}\,, \qquad\qquad \xb\= \bar{k}^+/\bar{p}^+\,.
\ee
These parameters are related to the usual momentum fractions the
gluons carry with respect to their parent proton, by
$x^{(\prime)}=(\xb\pm \xi)/(1\pm \xi)$.
The skewness is kinematically fixed to $\xi\simeq\xbj/2$ in a small $\xbj$
approximations. Hence, $x\neq x'$. This is to be contrasted with the
leading $\log(1/\xbj)$ approximation \ci{brodsky} where $x\simeq x'\simeq
\xbj$ is assumed and the GPD replaced by the usual gluon distribution
$g(x)$.

\begin{figure}[pt]
\includegraphics[width=.44\textwidth,bb=38 343 540 741,clip=true]
{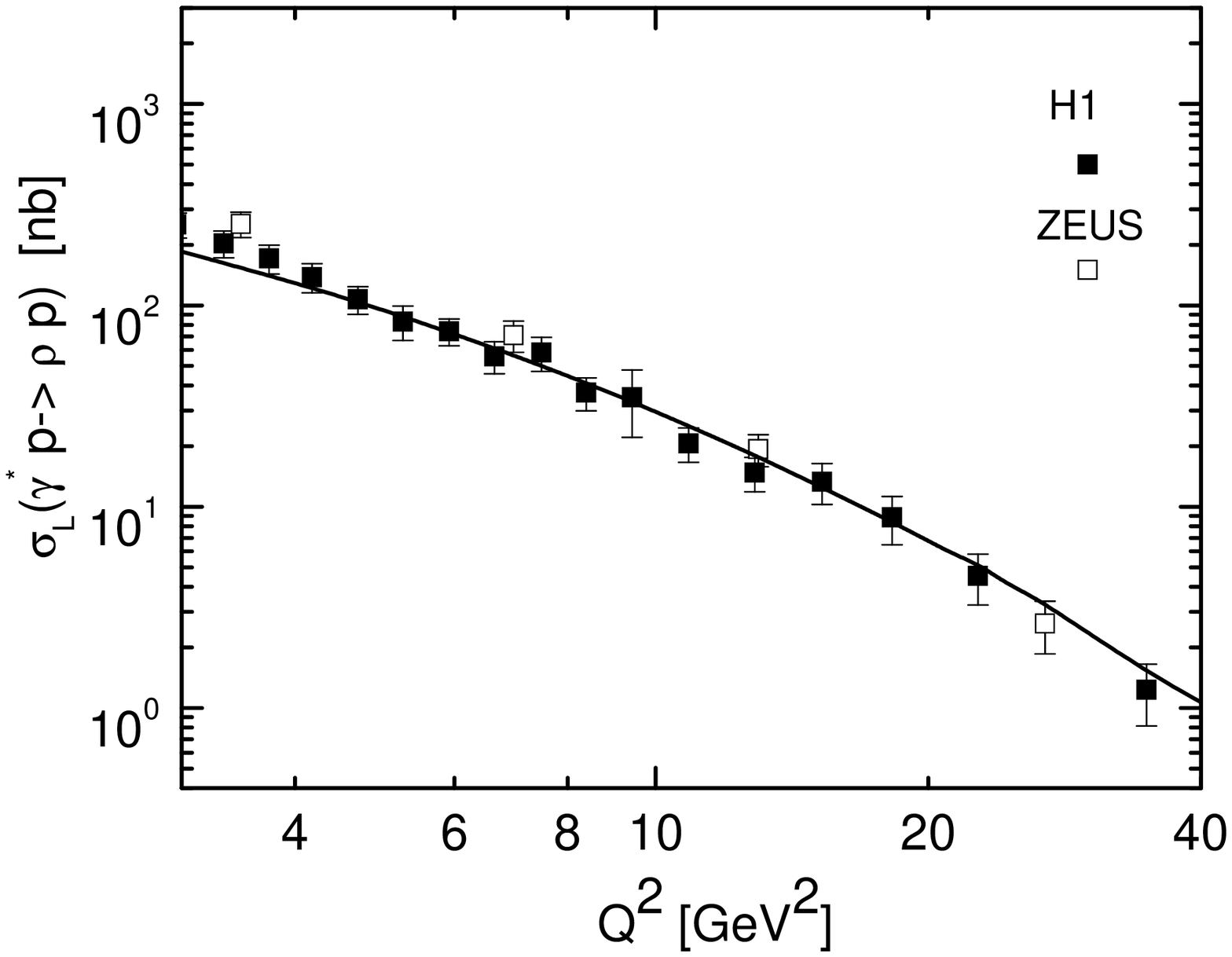}\hspace*{0.3cm}
\includegraphics[width=0.44\textwidth, bb= 49 352 530 733,clip=true]
{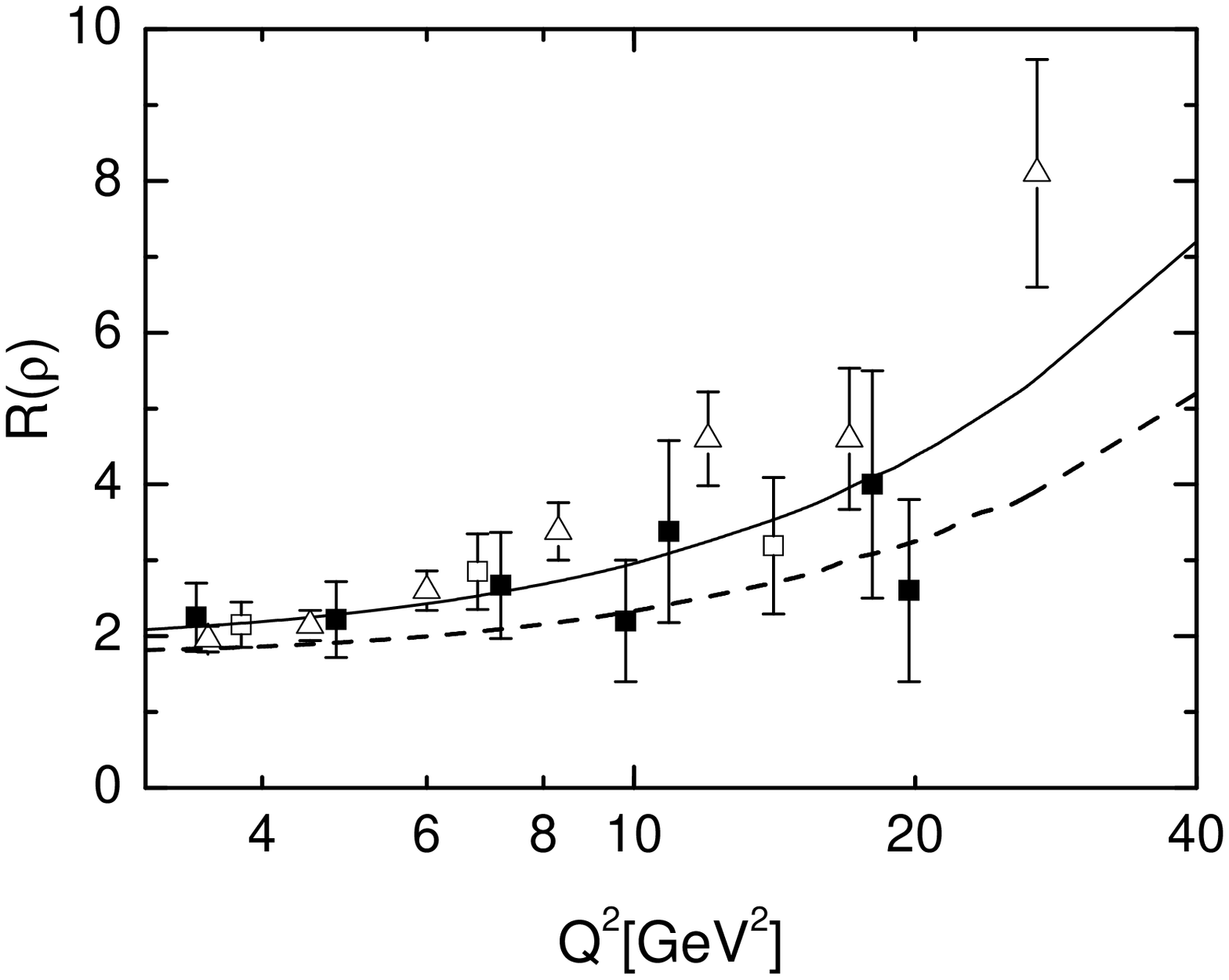}
\caption{Left: The integrated cross section for $\gamma_L p\to \rho p$
  versus $Q^2$ at $W\simeq 75\, \gev$. Right: The ratio of
  longitudinal and transverse cross sections for $\rho$ production 
  versus $Q^2$ at $W \simeq 75\, \gev$. Data taken from
 \protect\ci{h1} (filled squares) and \protect\ci{zeus} (open symbols). 
 The solid (dashed) lines are the results of Ref.\ \ci{golo} for the ratio 
 of differential (integrated) cross sections.} 
\label{fig:3}
\end{figure}

The handbag approach leads to the following proton helicity
non-flip amplitude
\be
M^V_{\mu' +,\mu +}\= \frac{e}{2} {\cal C}_V\, \int_0^1\,
                            \frac{d\xb}{(\xb+\xi)(\xb-\xi+i\varepsilon)}\,
                 \Big[H^V_{\mu' +,\mu +} + H^V_{\mu' -,\mu -}\Big]\,
                 H^g(\xb,\xi,t)\,.
\label{ampl}
\ee
Contributions from other GPDs can be neglected at small $\xbj$ and for
unpolarized protons.  
The photon and meson helicities are denoted by $\mu$ and $\mu'$,
respectively. The explicit labels in the full (subprocess) amplitude, $M^V$ ($H^V$)
refer to the helicities of the protons (gluons).\\
The GPDs are controlled by non-perturbative QCD. 
In the absence of an GPD analysis 
in analogy to those of the usual PDFs (see however
\ci{DFJK4}) one has to rely on a model. Its contruction is 
however not an easy matter since the GPDs are functions of three 
variables. Factorising the $t$ dependence from the $\xb, \xi$ one 
is probably incorrect. We therefore restrict ourselves to the 
forward direction and exploit the ansatz for a double distribution 
proposed in Ref.\ \ci{musa} ($n=1,2$)
\be
f(\beta,\alpha,t\simeq 0) \= g(\beta)\,
                 \frac{\Gamma(2n+2)}{2^{2n+1}\,\Gamma^2(n+1)}\,
               \frac{[(1-|\beta|)^2-\alpha^2]^n}{(1-|\beta|)^{2n+1}} \,.
\label{double-dis}
\ee 
The GPDs is then obtained by an integral over $f$
\be
H^{g}(\xb,\xi) = \Big[\,\Theta(0\leq \xb\leq \xi)
         \int_{x_3}^{x_1}\, d\beta + 
       \Theta(\xi\le \xb\leq 1) \int_{x_2}^{x_1}\, d\beta \,\Big] 
        \frac{\beta}{\xi}\,f(\beta,\alpha=\frac{\xb-\beta}{\xi})\,.
\label{gpd-model}
\ee
Using  the NLO CTEQ5M \ci{CTEQ} result as 
input we obtain the GPD $H^g$ shown in Fig.\ \ref{fig:1}.\\
The last item of the amplitude \req{ampl} to be discussed is the
subprocess amplitude. Its treatment is rather standard, it only
differs in detail from versions to be found in the literature
\ci{martin,frank}. In the modified perturbative approach invented by Sterman
and collaborators \ci{li92}, in which quark transverse momenta
are retained and gluonic radiative corrections in the form of a
Sudakov factor are taken into account, it reads
\be
{\cal H}^V\=\int \frac{d\tau d k^2_\perp}{\sqrt{2}16\pi^2} \Psi_V 
    \,{\rm Tr} \left\{(q\sla'+m_V)\eps_V\sla^*T_0 
       - \frac{k^2_\perp g_\perp^{\alpha\beta}}{2M_V}\, 
       \{(q\sla'+m_V)\eps_V\sla^*,\gamma_\alpha\}\, 
            \Delta T_\beta\right\}\,.
\label{amp-mpa}
\ee
Higher order terms in this expansion are not shown. Gaussians for the 
meson's wavefunctions, $\Psi_V=\Psi(\tau,k^2_\perp)$, are used which may depend on 
the polarization of the vector meson. There are two parameters 
specifying the wavefunction, the meson's decay constant and a
transverse size parameter. For longitudinally polarized vector
mesons these parameters are fairly well-known.
The first term in \req{amp-mpa} dominates for $V_L$ while it is
approximately zero for transversally polarized vector mesons ($V_T$). 
In the latter case the second term is 
the dominant one. Note that the soft physics parameter $M_V$ in this 
term is of order of the vector meson mass $m_V$. As can be seen from Eq.\
\req{amp-mpa} the $L\to L$ transition is dominant
while the $T\to T$ one is of relative order
$\langle k_\perp^2\rangle^{1/2}/Q$ and the $T\to L$ one of order
$\sqrt{-t}/Q$. The latter amplitude is tiny and only noticeable in
some of the SDMEs. All other transitions are negligible.
Eventual infrared singularities that may occur
for transitions to $V_T$, 
are regularized in the modified perturbative approach. \\
Before comparing the results to experiment I have to comment on the
$t$ dependence of the amplitudes. Exponentials in $t$ are assumed with
slopes $B^V_{LL(TT)}$ taken from experiment. Combined with the
calculated forward amplitudes one can evaluate the integrated cross 
sections and the SDME at small $t$. From \req{amp-mpa} one sees that
the size of the $T\to T$ amplitude is controlled by the following 
product of parameters 
\be
\Big| M_{TT}^V\Big| \propto \left(\frac{f_T^V}{M_V}\right)^2\,
\frac1{B^V_{TT}}\,.
\ee 
Without precise $t$-dependent data at disposal only this product is
probed. One can therefore, for instance, assume $B^V_{TT}\simeq
B^V_{LL}/2$. Combined with $M_V=m_V$ and $f^\rho_T=250\,\mev$ this
assumption provides reasonable results for vector meson
electroproduction, see Fig.\ \ref{fig:3}. An alternative choice is 
$B^V_{TT}\simeq B^V_{LL}$, $M_V=m_V$, $f_T^\rho=170\,\mev$ which leads
to practically the same results for the cross sections. Only the $t$ 
dependence of the SDME differs in both the cases. Given the accuracy 
of the present data \ci{h1,zeus} both the scenarios are in agreement
with experiment.\\  
In Fig.\ \ref{fig:3} the cross section $\sigma_L$ and the ratio
$R=\sigma_L/\sigma_T$ are displayed. The data on $R$
are extracted from the SDME measurements. This extraction is
problematic if the slopes are different. 
For comparison the ratio of the corresponding differential cross
sections is also shown in Fig.\ \ref{fig:3} (at $t\simeq -0.15\, \gev^2$). 
Results for the SDME of $\rho$ and $\phi$ mesons are also presented in
\ci{golo} in fair agreement with experiment.\\
Finally I want to comment on the $W$ dependence of the dominant
longitudinal cross section. It is given by the imaginary part of the
$L\to L$ amplitude with a correction of about $10\%$ from the real
part. The cross section is therefore approximately proportional to 
$|H^g(\xi,\xi)|^2$.
Through the model \req{gpd-model} the low-$x$ behaviour of the PDF 
$\xb g(\xb) \sim \xb^{-\delta (Q^2)}$ is tranferred to the GPD and one
finds
\be
\sigma_L \propto W^{-4\delta (Q^2)}\,.
\ee
The $Q^2$ dependence of $\delta$ is a consequence of evolution.  
Comparison with experiment reveals that this behaviour is in
remarkable agreement with the data within admittedly large errors. A
last remark: The expression for $\sigma_L$  
the GPD approach provides, is also obtained in the leading
$\log{1/\xbj}$ approximation given that the subprocess is treated
equally and that $H^g(\xi,\xi)$ is replaced by $2\xi g(2\xi)$. The
quality of this approximation is rather good for $\xi\lsim 10^{-2}$,
there is only an enhancement of the GPD by about $18\%$, the skewing
effect \ci{martin}. For increasing $\xi$ the approximation becomes gradually
worse.\\

I summarize: Vector meson electroproduction off unpolarized protons
at small $\xbj$ and small $t$ probes the GPD $H^g$. Calculating the
partonic subprocess within the modified perturbative approach (using
gaussian wavefunctions) fair agreement with HERA data on the
integrated cross sections for longitudinally and transversally
polarized virtual photons and the spin density matrix elements are
obtained for electroproduction of $\rho$ and $\phi$ mesons. It is to
be stressed that only the forward amplitudes are caluclated within the
GPD approach. Their $t$ dependencies are assumed to be exponentials with
slopes taken from experiment. The present data do, however, not fix
the slope of the $T\to T$ amplitude precisely. This treatment
of the $t$ dependence is unsatisfactory and improvements 
are required. In principle the GPD approach has the
potential to do better but the GPDs as a function
of $t$ are needed for that. It is also possible to go to larger values
of $\xbj$ with it. Some results on $\phi$ production at COMPASS
kinematics are presented in \ci{golo}.


\begin{thebibliography}{9}

\bibitem{rad96} A.V.\ Radyushkin, 
\emph{Phys.\ Lett.\ B {\textbf 385}, 333 (1996)};
J.C.\ Collins {\it et al.}, 
\emph{Phys.\ Rev.\ D} {\textbf 56}, 2982 (1997).

\bibitem{golo} S.~V.~Goloskokov and P.~Kroll,
  hep-ph/0501242, to be published in Eur.\ Phys.\ J.\ C.

\bibitem{piller} L.~Mankiewicz, G.~Piller and T.~Weigl,
\emph{Eur.\ Phys.\ J.\ C} {\textbf 5}, 119 (1998).

\bibitem{martin} A.~D.~Martin, M.~G.~Ryskin and T.~Teubner,
          \emph{Phys.\ Rev. D}\ {\textbf 62} (2000) 014022 

\bibitem{brodsky} S.~J.~Brodsky, {\it at al.},
\emph{ Phys.\ Rev.\ D} {\textbf 50}, 3134 (1994). 

\bibitem{DFJK4} M.~Diehl, T.~Feldmann, R.~Jakob and P.~Kroll,
\emph{Eur.\ Phys.\ J.\ C} {\textbf 39}, 1 (2005).

\bibitem{musa} I.~V.~Musatov and A.~V.~Radyushkin,
\emph{Phys.\ Rev.\ D} {\textbf 61}, 074027 (2000).

\bibitem{CTEQ} J.~Pumplin, D.~R.~Stump, J.~Huston, H.~L.~Lai,
  P.~Nadolsky and W.~K.~Tung, 
\emph{JHEP} {\textbf 0207}, 012 (2002).

\bibitem{frank} L.~Frankfurt, W.~Koepf and M.~Strikman,
        \emph{Phys.\ Rev.\ D} {\textbf 54}, 3194 (1996) 

\bibitem{li92} J.~Botts and G.~Sterman,
                   Nucl.~Phys.~B {\textbf 325}, 62 (1989).

\bibitem{h1} C.~Adloff {\it et al.}  [H1 Collaboration],
                \emph{Eur.\ Phys.\ J.\ {\textbf C13}, 371 (2000)}; 
S.~Aid {\it et al.}  [H1 Collaboration],
\emph{Nucl.\ Phys.\ B} {\textbf 468}, 3 (1996).

\bibitem{zeus} J.~Breitweg {\it et al.}  [ZEUS Collaboration],
\emph{Eur.\ Phys.\ J.\ C} {\textbf 6}, 603 (1999)
S.~Chekanov  [ZEUS Collaboration],
hep-ex/0504010.
\end{thebibliography}
\end{document}